# Elementary theory of earthquake source relaxation

A.V. Guglielmi

*Schmidt Institute of Physics of the Earth, Russian Academy of Sciences; Bol'shaya Gruzinskaya str., 10, bld. 1, Moscow, 123242 Russia; guglielmi@mail.ru*

*Abstract*: The elementary theory of relaxation of the source "cooling down" after the main shock of an earthquake is presented axiomatically. The names of the objects under study are given and the relationships between them are determined. A new basic concept of "earthquake source deactivation" is introduced and a procedure for calculating the deactivation coefficient from aftershock frequency measurement data is indicated. An important property of the system is that the axioms do not contain statements regarding the actual process of source relaxation. From two simple axioms a number of meaningful statements (theorems) logically follow. A two-stage mode of source relaxation was discovered. The sharp transition between stages has the character of a bifurcation. It is shown that the classical Omori law has limited applicability. It describes the evolution of aftershocks only at the first stage of relaxation. The well-known Hirano-Utsu law is not applicable to describe aftershocks either at the first or second stages of relaxation. The conclusions of the elementary theory allow for interesting generalizations that expand the possibilities of experimental and theoretical study of the source.

**Content**

## 1. Introduction

The source of an earthquake can be represented as a stressed-strained rock mass in a metastable state, in which a main discontinuity is formed spontaneously or induced, manifesting itself in the form of the main shock of the earthquake [Bolt, 1978]. After the main shock, the process of relaxation of the source begins. Figuratively speaking, the source "cools down". Relaxation is accompanied by the excitation of a large number of aftershocks at the source. Over time, the frequency of aftershocks decreases on average. According to Bath's law, the magnitude of aftershocks is less than the magnitude of the main shock by an amount exceeding $\Delta M = 1.1$ [Kasahara, 1981].

130 years ago, Fusakichi Omori summarized his observations of aftershocks and formulated the statement that the frequency of aftershocks decreases hyperbolically over time [Omori, 1894]. Chronologically, this was the first law of earthquake physics established by the inductive method of generalizing empirical data. The law is now widely known as Omori's law (see for example [Davison, 1930; Guglielmi, 2017]). Hirano and Utsu proposed an alternative theory for the evolution of aftershocks [Hirano, 1924; Utsu, 1961, 1962] (see also review [Utsu, Ogata, Matsu'ura, 1995]). According to their ideas, the frequency of aftershocks decreases over time according to a power law. In the literature, it is customary to call the power law the Utsu law [Salinas-Martínez, et al., 2023], the Omori-Utsu law [Rodrigo, 2021], or the modified Omori law [Ogata, Zhuang, 2006]. In my opinion, it should be called the Hirano-Utsu law. The Hirano-Utsu law underlies some developments in the theory of the evolution of aftershocks (see, for example, [Ogata, 1988; Ogata, Zhuang, 2006]).

The question of choosing between the Omori law and the Hirano-Utsu law is of some interest.

In this paper we will solve the issue using the deductive theory of source relaxation. We will see that a comparison of the theory with observations of aftershocks indicates the inapplicability of the Hirano-Utsu law to describe the evolution of aftershocks. As for Omori's law, it is applicable, but its applicability is limited only to the first stage of source relaxation.

The axiomatic method is interesting in itself, regardless of the above question. From the elementary axioms that we will present, a number of statements (theorems) follow that allow generalizations and, thereby, enrich the tools for the physical and mathematical study of aftershocks.

## 2. Axioms of the source relaxation theory

Let us formulate a system of axioms of the elementary theory of the earthquake source relaxation. The fundamental concept of the theory is the deactivation coefficient $\sigma(t)$ as a continuous function of time. An additional concept, borrowed from observational experience, is the frequency of aftershocks $n(t)$ as a differentiable function of time. We base the theory of source relaxation on two axioms:

1. There is a function of the source state $\sigma(t)$ called the deactivation coefficient.
2. The deactivation coefficient is calculated using formula

$$\sigma = -\frac{1}{n^2}\frac{dn}{dt},  \qquad (1)$$

in which $n(t)$ is the frequency of aftershocks.

We have called our deductive theory elementary in the sense that from the rich experience of empirical research on aftershocks and corresponding inductive inferences, we borrow only one concept, namely the concept of aftershock frequency. It is also necessary to clarify that our choice of formula (1) for calculating the deactivation coefficient is acceptable, but, generally speaking, the choice is ambiguous. The rationality of our choice will become clear in the next section of the paper, when we make sure that formula (1) allows us to easily visualize the hyperbolic decrease in the frequency of aftershocks over time.

We use axiomatics for the mathematical description of aftershocks in order to, if possible, give the theory the laconicism and style adopted in deductive science. From our axioms a number of simple statements (theorems) logically follow, the plausibility of which can be tested experimentally:

**1**. Using the auxiliary function $g(t)=1/n(t)$, the evolution of aftershocks is described by the simplest differential equation

$$\frac{dg}{dt}=\sigma(t). \qquad (2)$$

**2**. The inhomogeneous linear aftershock equation (2) is equivalent to the homogeneous nonlinear equation

$$\frac{dn}{dt}+\sigma n^{2}=0. \qquad (3)$$

**3**. The classical Omori law holds if and only if the deactivation coefficient is independent of time. Indeed, if $\sigma=\text{const}$, then the hyperbola is a solution to the evolution equation (3). If, on the other hand, the frequency of aftershocks decreases hyperbolically over time, then $\sigma=\text{const}$, as follows from formula (1).

**4**. The solution to the direct problem for the evolution equation (3) has the form

$$n(t)=\frac{n_{0}}{1+n_{0}\tau(t)}, \qquad (4)$$

where

$$\tau(t) = \int_0^t \sigma(t)\,dt. \qquad (5)$$

Here $n_0 = n(0)$ is the initial condition. Formula (4) is obtained by solving equation (3) using the method of separation of variables. It is worth paying attention to the fact that the frequency of aftershocks decreases hyperbolically if time is measured in units of proper time $\tau(t)$. The proper time in the source flows unevenly due to the non-stationary state of the source $\sigma(t)$.

**5**. The correct solution to the inverse problem for equation (3) has the form

$$\sigma(t) = \frac{d}{dt}\langle g(t) \rangle, \qquad (6)$$

where the angle brackets denote the optimal averaging of the auxiliary function. In fact, the trivial solution $\sigma = dg/dt$ is incorrect (unstable), since the original function $n(t)$ usually fluctuates quickly. Regularization of the problem in our case comes down to adequate smoothing of the auxiliary function.

## 3. Source relaxation theory in action

A number of previously unknown and very interesting results of experimental research on aftershocks, presented in reviews [Guglielmi, Klain, 2020; Zavyalov, Zotov, Guglielmi, Klein, 2022; Guglielmi, Klain, Zavyalov, Zotov, 2023], convincingly indicates the effectiveness of our deductive system. Here we present one result that seems extremely important. We will talk about a two-stage relaxation mode of the source. It turned out that the relaxation process is uneven. Relaxation of the source is clearly divided into two stages, and the transition from the first stage to the second occurs very abruptly.

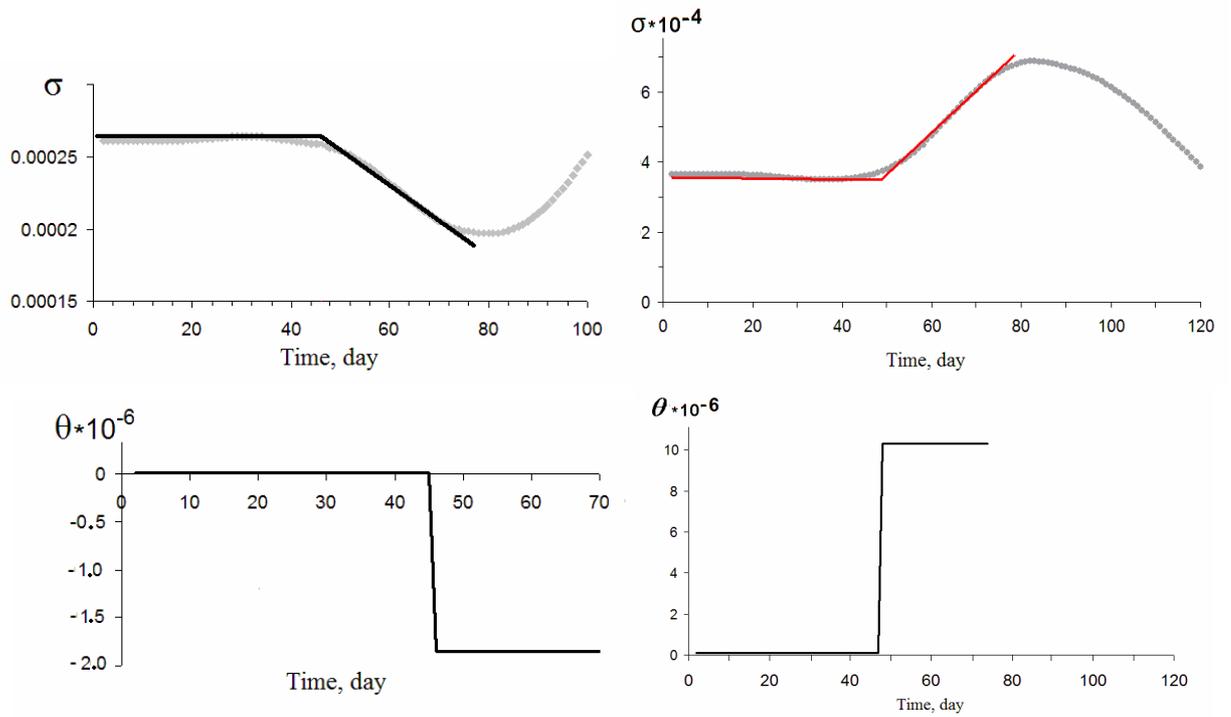

Illustration of the two-stage mode of aftershock evolution. The two left (right) panels show an event that occurred in Southern (Northern) California on 16.10.1999 (20.07.1986) with a main shock magnitude of $M = 7.1$ ($M = 5.9$). The straight line segments in the top row represent a piecewise linear approximation of the deactivation coefficient. The bottom row shows the function $\theta = d\sigma/dt$.

The two-stage relaxation mode is clearly visible in the figure, which is a collage of drawings borrowed from the works of [Guglielmi, Zotov, 2023; Guglielmi, Klain, Zavyalov, Zotov, 2023]. At the first stage, $\sigma = \text{const}$ and, thus, the classical Oyori law is strictly satisfied according to Theorem No. 3. The first stage of relaxation is naturally called the Omori epoch. The duration of the Omori epoch varies from case to case from several days to several months. The deactivation coefficient in the Omori epoch is lower, the higher the magnitude of the main shock. In the Omori epoch, the relaxation of the source proceeds in a completely predictable manner.

The beginning of the second stage is signaled by a jump in the derivative of the deactivation coefficient, reminiscent of a bifurcation (see figure). After bifurcation, the state of the source changes in a non-monotonically unpredictable manner. The search for a mechanism for the formation of a two-stage relaxation regime of the source is an urgent problem in earthquake physics.

## 4. Beyond elementary axioms

A real source is not a closed system. Its dynamics are influenced by geophysical fields, the sources of which are located outside the earthquake source. This type of impact often causes tremors and is usually called an exogenous trigger. Endogenous triggers are also known. Their sources are located inside the earthquake source (see below).

This statement is not an agreement (axiom) or a theorem. It does not claim to be a law of nature, but it should be taken into account when empirically studying the relaxation of an earthquake source. It is advisable to find a mathematical model of the trigger induction of tremors. The simplest model has the form of inhomogeneous differential equation

$$\frac{dn}{dt} + \sigma n^2 = f(t). \qquad (7)$$

We simply took the homogeneous evolution equation (3) and arbitrarily added a free term $f(t)$ to the right-hand side, simulating a trigger.

Let us point out here two endogenous triggers that are excited by the main shock [Guglielmi, Zotov, 2012, 2013; Guglielmi, Zotov, Zavyalov, 2014; Zotov, Zavyalov, Guglielmi, Lavrov, 2018]. One of them is pulsed and represents a round-the-world seismic echo of the main shock. The second trigger represents free vibrations of the Earth and is strictly periodic.

Equation (7) will become a stochastic differential equation if, instead of the deterministic function, we insert the Langevin source on the right side, i.e. delta-correlated random function with zero mean: $\langle f(t) \rangle = 0$, $\langle f(t) f(t') \rangle \propto \delta(t-t')$. In this case, equation (7) will simulate the impact of random noise on the source.

Adding a linear term to the right side of the nonlinear evolution equation (3) turns it into the well-known logistic Verhulst equation

$$\frac{dn}{dt} = n(\gamma - \sigma n). \qquad (8)$$

The meaning of this generalization is as follows [Guglielmi, Klain, 2024]. Elementary theory predicts that in asymptotics ($t \to \infty$) source relaxation leads to a stationary state with zero aftershock frequency. Generally speaking, this contradicts observations. After a sufficiently long time has elapsed after the main shock, a non-zero background of seismic activity is usually observed in the source. The logistic equation simulates this property of the source. The frequency of underground strikes in the background is $n_\infty = \gamma/\sigma$. For sufficiently strong main shocks, the inequality $n_0 \gg n_\infty$ holds. Moreover, in all cases known to us, $n \gg n_\infty$ at the first stage of relaxation. Thus, at the first stage, we can neglect the linear term in equation (8) and describe relaxation with good accuracy by equation (3), from which the Omori law follows, which is satisfied at the first stage according to observations.

Finally, another step that takes us beyond the limits of elementary theory is the formulation of the problem of the spatiotemporal evolution of aftershocks The problem is motivated by the experimentally discovered propagation of aftershock activity [Zotov, Zavyalov, Klain, 2018] (see also [Zavyalov, Zotov, Guglielmi, Klain, 2022; Guglielmi A.V., Zotov, 2024]). When interpreting observations, it was proposed to describe the evolution of aftershocks not by the ordinary differential equation (3), but by the Kolmogorov-Petrovsky-Piskunov partial differential equation, which has solutions in the form of nonlinear diffusion waves [Zotov, Zavyalov, Klain, 2018].

## 5. Discussion
### 5.1. Omori's law and Hirano-Utsu's law

130 years have passed since the first empirical law of earthquake physics was formulated. Fusakichi Omori found that after the main shock of an earthquake, the frequency of aftershocks decreases hyperbolically over time [Omori, 1894]. Omori's law is usually viewed as holistic, i.e. it is believed to describe the entire process of aftershock evolution. Meanwhile, a two-stage relaxation regime of the earthquake source was discovered. The transition from the first stage to the second occurs abruptly and resembles the phenomenon of bifurcation. It turned out that Omori's law is satisfied at the first stage of relaxation, but not at the second This important circumstance motivated our search for an axiomatic form of the theory of aftershocks.

Let us write Omori's hyperbolic law in its classical formulation [Onori, 1894]

$$n(t) = \frac{k}{c+t}. \tag{9}$$

Here time $t \geq 0$, and $k$ and $c$ are positive constants. According to formula (1), it follows that the deactivation coefficient does not depend on time. Thus, Omori's law is applicable to describe aftershocks, but its applicability is limited to the first stage of source relaxation.

The Hirano-Utsu law has the form

$$n(t) = \frac{k}{(c+t)^p}, \tag{10}$$

where $p > 0$ (see for example [Utsu, Ogata, Matsu'ura, 1995]). According to formula (1), it follows that the deactivation coefficient monotonically decreases over time at $p < 1$, and monotonically increases at $p > 1$. Thus, law (11) does not work at the first stage of source relaxation. But it does not work at the second stage either, since the real deactivation coefficient non-monotonically depends on time at the second stage.

### 5.2. State function of earthquake source

We introduced the deactivation coefficient into scientific circulation as a new variable of the earthquake source. But the deactivation coefficient only partially describes the state of the source, since the source begins to form long before the main shock. The main shock signals a catastrophic change in the state of the source, but the catastrophe (in the physical and mathematical sense of the word [Guglielmi, 2015]) is led by a long and complex tectonophysical process, during which conditions are formed for the inevitable formation of a main rupture in the continuity of rocks. The process of preparing the main shock is quite often, although not always accompanied by foreshocks. It is quite natural to try to use foreshocks to determine the state of the source during the preparation of the main shock. It should be said that the main shock is preceded not only by foreshocks, but also by a number of other precursors in the form of anomalous variations in geophysical fields. But in this work we will limit ourselves to foreshocks to form the function $s(t)$, which describes the states of the source both before and after the main shock.

The main rupture radically changes the state of the source, so it is reasonable to choose the function as a combination of two parts: $s = \{s_-, s_+\}$. Here $s_-$ ($s_+$) is a function of the state before (after) the main shock. It is clear that the $s_+$ coincides with the deactivation coefficient. This suggests the idea of using formula

$$s_\mp = \mp \frac{d}{dt}\langle g_\mp \rangle \qquad (11)$$

to calculate two branches of the state function based on measurement data of the foreshock frequency $n_-$ and the aftershock frequency $n_+$. Here $g_\mp = 1/n_\mp$.

Variation of functions $s_\mp(t)$ gives us additional information about the source as a dynamic system. The value of information increased after the discovery of two small but highly interesting categories of sources in which main shocks are excited, accompanied by an unusual combination of foreshocks and aftershocks [Zotov, Guglielmi, 2021; Guglielmi, Zotov, 2024]. These categories were called mirror and symmetric earthquake triads. In the mirror triad, the average frequency of foreshocks is higher than the average frequency of aftershocks, and aftershocks may be absent altogether. It is important to note that for foreshocks in the mirror triad, Zotov's law is satisfied, similar to Bath's law for aftershocks, mentioned in the Introduction [Zotov, Guglielmi, 2021]

In a symmetric triad, the average frequencies of foreshocks and aftershocks are the same. We present here information about the variation of $s_\mp(t)$ upon excitation of symmetric triads. It turned out that

$$s_{\mp}(t) = s_0 \exp(-at^2), \tag{12}$$

We know that in the classical triad $s_+ = \text{const}$ at the first stage of source relaxation. Thus, in a symmetric triad the property $s_+ = \text{const}$ is sharply violated. An interesting task arises of finding the specific tectonophysical conditions for the formation of sources that give rise to symmetrical triads. There is hope that solving the problem will shed light on the origin of the first stage of relaxation of the source that excite classical triads.

## 6. Conclusion

In this paper, we outlined the elementary theory of relaxation of the source "cooling down" after the main shock of an earthquake. We have determined the names of the objects under study and formulated axioms that must govern the relationships between them. From two simple axioms a number of meaningful statements (theorems) logically follow. The axiomatics provided the rationale for a new method of experimental research, which made it possible to establish a two-stage mode of source relaxation. It was found that the classical Omori law is satisfied for a limited period of time during the relaxation of the source. Thus, the system of axioms made it possible to logically describe the law known for 130 years and finally determine the scope of its applicability. A generalization of the concept of deactivation, which underlies the axiomatic theory, gave an idea of the existence of a state function that describes the dynamics of the source not only after, but also before the main shock of the earthquake.

*Acknowledgments*. I express my deep gratitude to B.I. Klain, A.D. Zavyalov and O.D. Zotov for discussions of problems in earthquake physics. I sincerely thank A.L. Buchachenko, F.Z. Feygin and A.S. Potapov for their interest in the work and support. The work was carried out according to the plan of state assignments of Schmidt Institute of Physics of the Earth, Russian Academy of Sciences.